\documentclass[11pt]{article}

\usepackage{latexsym,amsfonts,amsmath,amssymb,wasysym,stmaryrd,enumerate,epsfig}
\usepackage{mathrsfs}
\usepackage{theorem}

\numberwithin{equation}{section}
\newtheorem{definition}{Definition}[section]

\newtheorem{theorem}[definition]{Theorem}

\newcommand{\be}{\begin{equation}}
\newcommand{\ee}{\end{equation}}
\newcommand{\beu}{\begin{equation*}}
\newcommand{\eeu}{\end{equation*}}
\newcommand{\bea}{\begin{eqnarray}}
\newcommand{\eea}{\end{eqnarray}}
\newcommand{\beaa}{\begin{eqnarray*}}
\newcommand{\eeaa}{\end{eqnarray*}}
\newcommand{\bmx}{\begin{pmatrix}}
\newcommand{\emx}{\end{pmatrix}}
\newcommand{\Dih}{{\mathrm{Dih}}}

\newcommand{\del}{\partial}

\newcommand{\inZm}{\in \mathbb Z_m}

\renewcommand{\k}{\, b}
\newcommand{\q}{\, a}
\renewcommand{\d}{\, d}
\newcommand{\e}{\, e}
\newcommand{\dd}{\, D}

\newcommand{\proof}{{\noindent\textbf{Proof. }}}
\newcommand{\finproof}{{\hfill \rule{5pt}{5pt}}}

\newcommand{\half}{\frac{1}{2}}

\newcommand{\nn}{\nonumber}

\newcommand{\K}{{\mathscr K}}

\newcommand{\eps}{\epsilon}

\newcommand{\Ad}{{\rm Ad}}

\newcommand{\Z}{{\mathbb Z}}

\renewcommand{\P}{{\mathscr P}}

\newcommand{\Q}{{\mathscr Q}}

\newcommand{\cH}{{\cal{H}}}
\newcommand{\Is}{I_{\text{spin}}}

\advance\textheight by \topskip
\begin{document}

\baselineskip 17pt
\parindent 18pt
\parskip 9pt

\begin{flushright}
\break

DCPT-07/41

\end{flushright}
\vspace{1cm}
\begin{center}
{\LARGE {\bf Sutherland Models  }}

{\LARGE {\bf for Complex Reflection Groups}}\\[4mm]
\vspace{1.5cm}
{\large  N. Cramp\'e\footnote{
crampe@sissa.it}$^{a,b}$ and C. A. S. Young\footnote{charlesyoung@cantab.net}$^c$}
\\
\vspace{10mm}
{ \emph{$^a$ International School for Advanced Studies,\\
Via Beirut 2-4, 34014 Trieste, Italy}}

{ \emph{$^b$ Istituto Nazionale di Fisica Nucleare\\
Sezione di Trieste}}

{ \emph{$^c$ Department of Mathematical Sciences, University of Durham\\
South Road, Durham DH1 3LE, UK}}

\end{center}

\vskip 1in
 \centerline{\small\bf ABSTRACT}
\centerline{
\parbox[t]{5in}{\small There are known to be integrable Sutherland models associated to every real
root system -- or, which is almost equivalent, to every real reflection group. 
Real reflection groups are special cases of \emph{complex} reflection groups. In this paper we
associate certain integrable Sutherland models to the classical family of complex reflection
groups. Internal degrees of freedom are introduced, defining dynamical spin chains, and the freezing
limit taken to obtain static chains of Haldane-Shastry type. By considering the relation of these
models to the usual $BC_N$ case, we are led to systems with both real and complex reflection groups
as symmetries. We demonstrate their integrability by means of new Dunkl operators, associated to
wreath products of dihedral groups.}}

\vspace{1cm}

\newpage

\section{Introduction}

\begin{figure}
\begin{center}
\epsfig{file=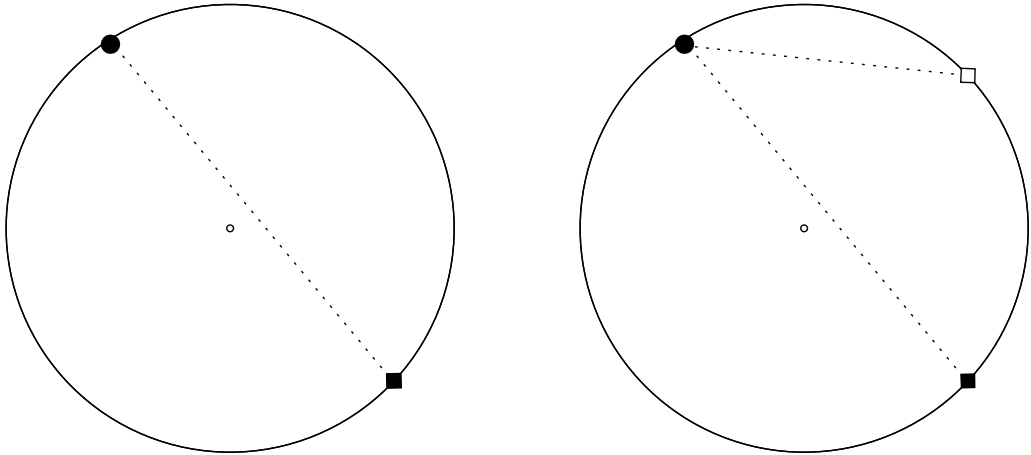,height=6cm}
\caption{Interactions between two particles ($\bullet$ and $\blacklozenge$) in the Sutherland models
associated to the $A$ (left) and $D$ (right) series of real reflection groups. The image of
$\blacklozenge$ under reflection is drawn as $\lozenge$.}
\label{fig1}
\end{center}
\end{figure}

The Sutherland model \cite{S} is an important and much-studied integrable quantum-mechanical system.
It describes $N$ particles moving on a circle, whose pairwise interactions are determined
by a potential proportional to the inverse square of the chord-length separating the particles (as
in figure \ref{fig1}). The
Hamiltonian, in the simplest case of identical spinless bosons, is
\be H = -\half \sum_{i=1}^N \frac{\del^2}{\del x_i^2} 
     + \lambda \sum_{i\neq j} \frac{1}{\sin^2\left( \half \left(x_i - x_j\right) \right)}
.\label{S1}\ee 
The model was first introduced in \cite{Sutherland}. It, and the wider family of
Calogero-Sutherland-Moser models \cite{Calogero} to which it belongs, have since appeared in areas
physics apparently far removed from the original condensed-matter context: see for example
\cite{HEP}. Operator methods were used to solve the system in \cite{brink,LV}; the Yangian symmetry
of the model was derived in \cite{haha,BGHP}. For recent reviews and references to the extensive
literature see \cite{rev,etingrev,PolyRev}.

The Hamiltonian (\ref{S1}) is closely related  to the Coxeter group $A_{N-1}$, because the potential
can be written as $\sum_{\alpha\in\Delta}\sin^{-2}(\half( \mathbf x \cdot \pmb \alpha))$, where
$\Delta = \{ \pmb\eps_i-\pmb\eps_j : i\neq j\}$ is the root system of $A_{N-1}$. Similar integrable
models exist also for all other finite Coxeter groups \cite{OPq,KKS,OPreps,BCS,BST,E6E7}. The
classical
families $BC_N$ and $D_N$ describe systems with boundaries, via a kind of method of images: in the
case of $D_{N}$, whose roots are $\pm \pmb \eps_i
\pm\pmb\eps_j$, one has
\be H = -\half \sum_{i=1}^N \frac{\del^2}{\del x_i^2} 
     + \lambda \sum_{i\neq j} \left(\frac{1}{\sin^2\left( \half \left(x_i - x_j\right) \right)}+ 
               \frac{1}{\sin^2\left( \half \left(x_i + x_j\right)\right) }\right),\label{SD}\ee 
and the second term describes the interaction of particle $i$ with the image of particle $j$ in the
boundary (see figure \ref{fig1}). In the $B$ and $C$ cases, the extra roots $\propto\pm\pmb\eps_i$
give an
interaction between the particles and the boundary.

\begin{figure}
\begin{center}
\epsfig{file=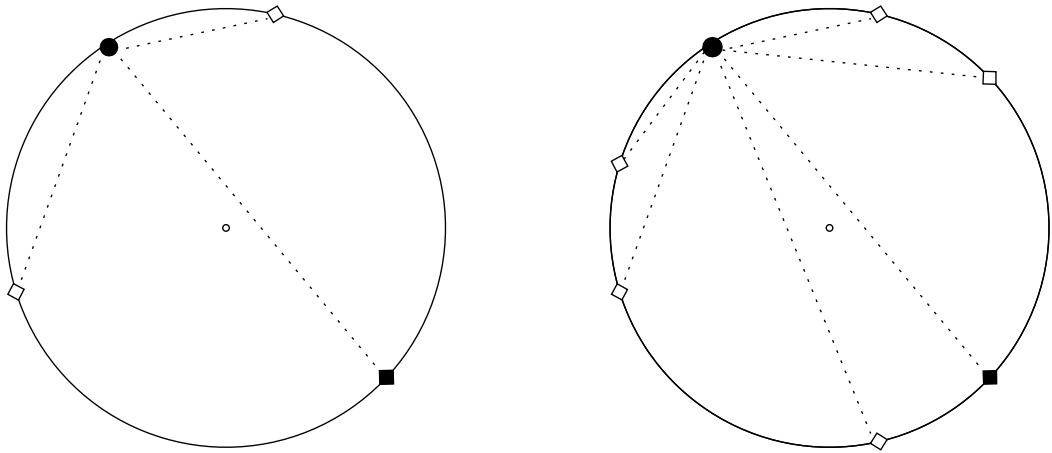,height=6cm}
\caption{Interactions between particles $\bullet$ and $\blacklozenge$ for the Sutherland models
associated to the classical complex reflection group $G(3,1,N)$  (left) and a wreath product of a
dihedral group (right). The images of $\blacklozenge$ are drawn as $\lozenge$.}  
\label{fig2}
\end{center}
\end{figure}

These families, $ABCD$, of course exhaust the classical irreducible finite Coxeter groups. Finite
Coxeter groups are finite real reflection groups \cite{Humphreys}: that is, subgroups of the
orthogonal group generated by a finite number of elements $s\in O(N)$ such that 
\be s^2=1, \qquad\text{$s$ has eigenvalue $+1$ with multiplicity $N-1$.} \label{realrefl}\ee
But it is possible to weaken the first of these
requirements and consider subgroups of $U(N)$ generated by finitely many $s\in U(N)$ obeying  
\be s^{n(s)}=1,\qquad\text{$s$ has eigenvalue $+1$ with multiplicity $N-1$} \label{comprefl} \ee
for some $n(s)\in\mathbb N$. By doing so one obtains \emph{complex reflection groups}. The
irreducible finite complex reflection groups were classified in \cite{ST}. There is one `classical'
three-parameter family $G(pr,p,N)$, $p,r,N\in \mathbb N$, which includes the four classical families
of real reflection groups as special cases, and then 34 exceptional cases. 

In this paper our main goal is to construct Sutherland models for classical complex reflection
groups. At first sight it is perhaps not clear that one should expect this to work,
because complex reflection groups lack a great deal of the usual structure that comes with real
reflection groups. The notions of root system, length function and  Coxeter graph are either absent
or, at best, less natural in the complex case \cite{CR} -- and the definition of the
$BCD$ Sutherland models sketched above appears to rely explicitly on the root system data. 
Nevertheless, it turns out that there do exist integrable models of Sutherland type associated to
complex reflection groups in a very natural fashion.  The basic idea is sketched in figure
\ref{fig2} -- each particle has a number of images, but now these images are generated by rotations.

The plan of this paper is as follows. In section \ref{algebra} we begin with some algebraic
preliminaries on complex reflection groups, and then introduce a key tool, Dunkl operators
\cite{duop}, from which we construct integrable Hamiltonians of Sutherland type. These models turn out to be members of a class of Calogero-Sutherland models first introduced and solved in \cite{P1998nm,P1998gg}. 
In section
\ref{sec:spin} we introduce models with internal ``spin'' degrees of freedom, and in section
\ref{sec:static} static or ``frozen'' chains in which the spins are in fact the only degrees of
freedom.  

In section \ref{dihed} we turn to models in which the set of images of each particle is generated by
a dihedral group, as illustrated in figure \ref{fig2}. As we shall see, these models possess both a
complex reflection group and a real reflection group as symmetries, embedded within a larger group
which will turn out to be a wreath product of a dihedral group. An important part of our
construction will be the introduction of new Dunkl operators, associated to such wreath products. We
then go on to introduce spin degrees of freedom and static chains with dihedral symmetry. 

We conclude by noting some open questions -- primarily of solution and Hamiltonian reduction --
concerning the new models of this paper, and some broader reasons for investigating integrable
systems with complex reflection groups as symmetries. 

\section{Complex Reflection Algebras}
\label{algebra}

\subsubsection*{Classical complex reflection groups}

The complex reflection group $G(m,1,N)$ is generated by 
$\{\q, \e_1,\e_2, \dots ,\e_{N-1}\}$, subject to the relations
\bea \e_i^2 = 1\quad & \quad \e_i \e_{i+1} \e_i = \e_{i+1} \e_i \e_{i+1}\quad& 
                     \quad \e_i \e_j = \e_j \e_i \quad (|i-j|>2) \nn\\
   \q^m = 1 \quad& \quad \q \e_1 \q \e_1 = \e_1 \q \e_1 \q \quad& \quad \q \e_j = \e_j \q \quad
(j>1)\label{gm1N}\eea 
The $\e_i$ generate a copy of the permutation group $S_N$ on $N$ objects, $\q$ generates a copy of
$\mathbb Z_m = \mathbb Z/m \mathbb Z$, and the full group is a semidirect product $G(m,1,N)=
\left(\mathbb Z_m\right)^N\rtimes S_N$. This structure is sometimes referred to as the \emph{wreath
product} of $\mathbb Z_m$ with $S_N$, denoted $\mathbb Z_m \wr S_N$. It will be convenient to write
\be \P_{ij} = \P_{ji} = \e_i \e_{i+1} \dots \e_{j-1} \dots \e_{i+1} \e_i\quad (i<j)\label{Pdef}\ee 
for the transposition $i\leftrightarrow j$ (in particular $\e_i = \P_{i i+1}$) and 
\bea \Q_1 &=& \q \\\Q_i &=& \P_{i1} \Q_1 \P_{i1} \quad (i>1).\label{Qdef}\eea
In terms of these elements the defining relations imply, and can be recovered from,
\bea \P_{ij}^2=1\quad&\quad\P_{ij}\P_{jk}=\P_{ik}\P_{ij} = \P_{jk}\P_{ik} 
\quad&\quad \P_{ij}\P_{kl}=\P_{kl}\P_{ij}\nn\\
&\P_{ij}\Q_i = \Q_j \P_{ij}\quad&\quad \P_{ij}\Q_k=\Q_k\P_{ij} \nn\\
\Q_i^m=1\quad&\quad \Q_i \Q_j = \Q_j \Q_i&  \qquad\qquad (i,j,k,l \text{ all distinct}).
\label{Gm1N}\eea

For any divisor $p$ of $m$, the complex reflection group $G(m,p,N)$ is the subgroup of $G(m,1,N)$
generated by \be\{\q ^p, \q ^{-1}\e_1 \q , \e_1, \e_2, \dots ,\e_{N-1}\}.\label{gmpN} \ee
The classical \emph{real} reflection groups occur as the special cases $A_{N-1} = S_N = G(1,1,N)$,
$BC_N = G(2,1,N)$ and $D_{N} = G(2,2,N)$.\footnote{But note that, in what follows, the models we
construct do not reduce, in the $BC$ and $D$ cases, to the standard Sutherland Hamiltonians, \`a la
(\ref{SD}), for these groups. We return to this point in section \ref{dihed}.}

\subsubsection*{Extended degenerate affine Hecke algebras}

We will overload notation slightly by using $G(m,1,N)$ also to refer to the group 
algebra of $G(m,1,N)$ over $\mathbb C$. Let us define $H_\lambda(m,1,N)$, 
$\lambda \in \mathbb C$, to be the algebra generated by 
\be
\{\q , \d, \e_1, \e_2,\dots, \e_{N-1}\},
\ee 
obeying (\ref{gm1N}) and the further relations 
\be \q  \d=\d \q \qquad\qquad \d \e_1 \q  \e_1 = \e_1 \q  \e_1 \d \qquad\qquad \e_j \d = \d \e_j
\quad (j>1)\label{drels1}\ee
\be \d \e_1 \d \e_1 + \lambda \d\sum_{s\inZm} \q ^{s} \e_1 \q ^{-s} = \e_1 \d \e_1 \d +
\lambda \sum_{s\inZm} \q ^{s} \e_1 \q ^{-s} \d \, .\label{drels}\ee
In addition to the $\Q_i$ and $\P_{ij}$ of (\ref{Pdef}--\ref{Qdef}), it is also useful to introduce
$d_1,\dots ,d_N$, defined recursively by
\bea d_1 &=& \d\\ d_{i+1} &=& \P_{ii+1} d_i \P_{ii+1} 
 + \lambda \sum_{s \inZm}\Q_i^{s} \P_{ii+1} \Q_i^{-s} 
\quad (i=1,\dots,N-1).\label{dprops}\eea
It follows from (\ref{drels1}--\ref{drels}) that 
\be \left[ \,d_i, d_j \right] = 0, \quad \quad \left[\,d_i, \Q_j\right] = 0\label{dprops2}\ee 

$H_\lambda(m,1,N)$ can be regarded as an affinization of $G(m,1,N)$, with $d$ in a very loose sense
a ``lowest root''. Indeed, when $m=1$ the sums
above collapse and one recovers the relations $\P_{ii+1} d_i = d_{i+1} P_{ii+1} + \lambda$ of the
degenerate affine Hecke algebra, first introduced in \cite{AffineHecke}.

We may define also $H_\lambda(pr,p,N)$, an affinization of $G(pr,p,N)$, to be the subalgebra of
$H_\lambda(pr,1,N)$ generated by $\{\q ^p, \d, \q ^{-1} \e_1 \q , \e_1, \e_2,\dots ,\e_{N-1}\}$.
Note that the relation (\ref{drels}) does not conflict with closure, because $G(pr,p,N)$ does
contain all the elements
\be \q ^{-s} \e_1 \q ^s =  \e_1 \left(\e_1 \q ^{-1} \e_1 \q \right)^{s} .\ee
Extended degenerate affine Hecke algebra associated with the $BC_N$ reflection groups appeared
previously in \cite{caduk,F02}. They differ from definition of $H_\lambda(2,1,N)$ here though: in
particular, the relation $[d_i,\Q_j]=0$ does not hold there. 

\section{Dunkl operators and Hamiltonians}

\subsubsection*{Realization of $H_\lambda(m,1,N)$}

The next stage is to realize these abstract algebraic relations in a concrete physical model.
Consider a quantum-mechanical system of $N$ particles on the unit circle. Let $q_i=\exp(ix_i)$ be
the position operator of the $i^{th}$ particle and write the position-space wavefunction as 
\be \psi(q_1,q_2,\dots, q_N).\ee 
Let $\P_{ij}$ be the operator which transposes the positions of particles $i$ and $j$,
\be \P_{ij} \psi(\dots,q_i,\dots, q_{j},\dots) = \psi(\dots,q_{j},\dots, q_i,\dots)\ ,\ee 
and $\Q_i$ the operator which rotates particle $i$ through $(\frac{1}{m})^{th}$ of a revolution,
\be \Q_{i} \psi(\dots,q_i,\dots) = \psi(\dots,\tau
q_{i}\dots),\quad\text{where}\quad\tau=\exp\left({2\pi i \over m}\right) \ee 
It is easy to see that these $\P_{ij}$ and $\Q_i$ satisfy the defining relations (\ref{Gm1N}) of
$G(m,1,N)$.
We also have that
\bea &&\Q_i\ q_i =\tau\ q_i\ \Q_i\qquad \quad \Q_i\ \frac{\del}{\del q_i} =\tau^{-1}\
\frac{\del}{\del q_i}\ \Q_i \\
&&\P_{ij} q_i=q_{j} \P_{ij} \qquad \quad \P_{ij}\frac{\del}{\del q_i} = \frac{\del}{\del
q_j}\P_{ij}.
\eea
The crucial step is the introduction of differential operators, called Dunkl operators \cite{D,
Poly1,brink}, that realize the algebraic relations of the $d_i$. The problem of
finding such operators for complex reflection groups was solved in \cite{duop}. Following that
paper,  with minor modifications that will allow us to obtain a slightly more elegant Hamiltonian,
we define
\bea
d_i&=&q_i \frac{\del}{\del q_i} +\lambda\sum_{j\neq i}\sum_{s\in\mathbb Z_m} \frac{q_i}{q_i-\tau^s
q_j} \Q_i^{-s}\P_{ij}\Q_i^s-\lambda\sum_{j>i}\sum_{s\in\mathbb
Z_m}\Q_i^{-s}\P_{ij}\Q_i^s\label{Dunkl}\\
&=&q_i \frac{\del}{\del q_i} +\lambda\sum_{j< i}\sum_{s\in\mathbb Z_m}\frac{q_i}{q_i-\tau^s q_j}
\Q_i^{-s}\P_{ij}\Q_i^s + \lambda\sum_{j> i}\sum_{s\in\mathbb Z_m}\frac{\tau^s q_j}{q_i-\tau^s q_j}
\Q_i^{-s}\P_{ij}\Q_i^s.\eea

\begin{theorem}\label{th1}
These provide a realisation of $H_\lambda(m,1,N)$.
\end{theorem}

This is essentially theorem (3.8) of \cite{duop}, and the same strategy of proof works here. But
when we come to introduce new Dunkl operators for wreath products of dihedral groups, in section
\ref{dihed}, it will be useful to have noted the following alternative

\proof
It is easy to verify that (\ref{dprops}) holds: were it not for the final term on the right
of (\ref{Dunkl}), the $\d_i$ would obey $\P_{ij} \d_i \P_{ij} = \d_j$. The final term involves an
ordering of the particles and is responsible for the extra piece in (\ref{dprops}). 

It remains to show that $\d_1 = \d$ obeys (\ref{drels1}) and (\ref{drels}). The first of these is
straightforward to check by direct computation\footnote{We sketch the arguments, for the more
involved case of dihedral groups to be considered in section \ref{dihed}, in an appendix.}. The
second is nothing but the statement that the  Dunkl operators commute:
\be [\d_1, \d_2 ] = 0,\ee
which is really the key property. To prove it, first recall the Dunkl operators of the $A_{N-1}$
case:
\be Z_i = q_i\frac{\del}{\del q_i} + m\lambda\sum_{j\neq i}\frac{q_i}{q_i-q_j} \P_{ij} 
                                - m\lambda\sum_{j>i} \P_{ij} \ee
where we have chosen the coupling to be $m\lambda$. Observe that then
\be \d_i = \frac{1}{m}\sum_{s \inZm} \Q_i^{-s} Z_i \Q_i^s.\ee
This motivates the definition of a family of projectors: we write
\be \Ad(\Q) (X) = \Q^{-1} X \Q \ee
and define
\be \Pi^r_i = \frac{1}{m} \sum_{s \inZm} \tau^{sr} \Ad(\Q_i^s)\label{Pri}.\ee 
These obey \be id=\sum_{r\inZm} \Pi^r, \qquad\qquad\Pi^r \Pi^t = \delta^{r,t} \Pi^r.\ee 
Also, for any $A$ and $B$, 
\be\Q \left(\Pi^r A\right) \left(\Pi^t B\right) = \tau^{r} \left(\Pi^r A\right) \Q \left(\Pi^t
B\right) 
     = \tau^{r+t} \left(\Pi^r A\right) \left(\Pi^t B\right) \Q,\ee  
so $\Pi^{r+t} \left(\Pi^r A\right) \left(\Pi^t B\right) =\left(\Pi^r A\right) \left(\Pi^t B\right)$.
It follows that
\be \Pi^0 AB = \sum_{r,t\in\mathbb Z_m} \Pi^0  \left(\Pi^r A\right) \left( \Pi^{t} B\right) =
\sum_{r\in\mathbb Z_m} \left(\Pi^r A\right) \left( \Pi^{-r} B\right).\ee

Armed with these facts we argue as follows. Given the result \cite{D,Poly1,brink} that
\be \left[ Z_i , Z_j \right] = 0 \ee
we have in particular that
\be 0= \Pi_i^0 \Pi_j^0 \left[ Z_i , Z_j \right] =
        \sum_{r,t\inZm} \left[ \Pi_i^r \Pi_j^t Z_i, \Pi_j^{-t} \Pi_i^{-r} Z_j\right].\label{ppzz}\ee
But one may compute, for all $i\neq j$,
\bea \Pi_i^r \Pi_j^t Z_i &=& \delta^{r,0} \delta^{t,0} q_i \frac{\del}{\del q_i} + 
 \delta^{t,0} \lambda \sum_{s\inZm} \tau^{rs} \left( \sum_{h\not\in\{i,j\}} \frac{q_i}{q_i-\tau^s
q_h} \Q_i^{-s} \P_{ih} \Q_i^s - \sum_{h>i, h\neq j} \Q_i^{-s} \P_{ih} \Q_i^s \right) \nn\\
&&{} + \delta^{r+t,0}\lambda\sum_{s\inZm} \tau^{rs}\left( \frac{q_i}{q_i-\tau^s q_j} \Q_i^{-s}
\P_{ij} \Q_i^s - \theta^{j>i} \Q_i^{-s} \P_{ij} \Q_i^s \right).\label{ppz}\eea
The only terms in (\ref{ppzz}) which can survive in view of the $\delta$'s here are 
\bea 0&=& \left[ \Pi_i^0 \Pi_j^0 Z_i , \Pi_i^0 \Pi_j^0 Z_j \right] 
   + \sum_{t\inZm, t\neq 0}\left[\Pi_i^t \Pi_j^{-t} Z_i, \Pi_j^{t} \Pi_i^{-t} Z_j\right] .\eea
The second term (with $i>j$, without loss of generality) is
\bea &&  
 \sum_{t\inZm, t\neq 0} \left[ \sum_{s\inZm } \tau^{ts} \frac{q_i}{q_i-\tau^s q_j} \Q_i^{-s} \P_{ij}
\Q_i^s ,
 \sum_{s'\inZm} \tau^{ts'} \left( \frac{q_j}{q_j - \tau^{s'} q_i} - 1 \right) \Q_j^{-s'} \P_{ji}
\Q_j^{s'}\right] \nn\\
&=& - \sum_{t\inZm, t\neq 0} 
\left[  \sum_{s\inZm } \tau^{ts} \frac{q_i}{q_i-\tau^s q_j} \Q_i^{-s} \P_{ij} \Q_i^s, 
\sum_{s'\inZm} \tau^{-ts'}\frac{q_i}{q_i-\tau^{s'} q_j}\Q_i^{-s'} \P_{ij} \Q_i^{s'}\right],\eea
and this vanishes. (In the sum over $t$, the summands cancel in conjugate pairs, $t$ against $-t$,
except for the possible $\tau^t=-1$ term which is zero by itself.) But now since $\Pi_i^0 \Pi_j^0
Z_i=\d_i$ by (\ref{ppz}), we have that indeed 
\be 0= \left[d_i,d_j\right],\ee
completing the proof.
\finproof

\subsubsection*{Hamiltonians}
It follows from the discussion above that the quantities 
\be I^{(k)} = \sum_{i=1}^N d_i^k \label{Hi}\ee
form a commuting set. The $I^{(k)}$ are algebraically independent for $k=1,2,\dots N$, and these
give $N$ commuting conserved quantities of the model with Hamiltonian 
\bea
\label{ham}
 H= I^{(2)} &=& \sum_{i=1}^N \left(q_i\frac{\del}{\del q_i}\right)^2
-2\lambda \sum_{i<j}\sum_{s\in\Z_m}\frac{\tau^s q_iq_j}{(q_i-\tau^s q_j)^2}
(\lambda+\Q_i^{-s}\P_{ij}\Q_i^s)\\
&=& \sum_{i=1}^N \left(q_i\frac{\del}{\del q_i}\right)^2
-\lambda \sum_{i\neq j}\sum_{s\in\Z_m}\frac{\tau^s q_iq_j}{(q_i-\tau^s q_j)^2}
(\lambda+\Q_i^{-s}\P_{ij}\Q_i^s)
\eea
which is therefore, by construction, integrable. After the change of coordinates $q_j= \exp(ix_j)$
the Hamiltonian takes the form
\be
\label{ham-sin}
H=-\sum_{i=1}^N \frac{\del^2}{\del x_i^2}
+\frac{\lambda}{4} \sum_{i\neq j}\sum_{s\in\mathbb Z_m}\
\frac{1}{\sin^2\left(\half\left(x_i-x_j+{2\pi s\over m}\right)\right)}
\left(\lambda+\Q_i^{-s}\P_{ij}\Q_i^s\right)
\ee

\begin{figure}[ht]
\begin{center}
\epsfig{file=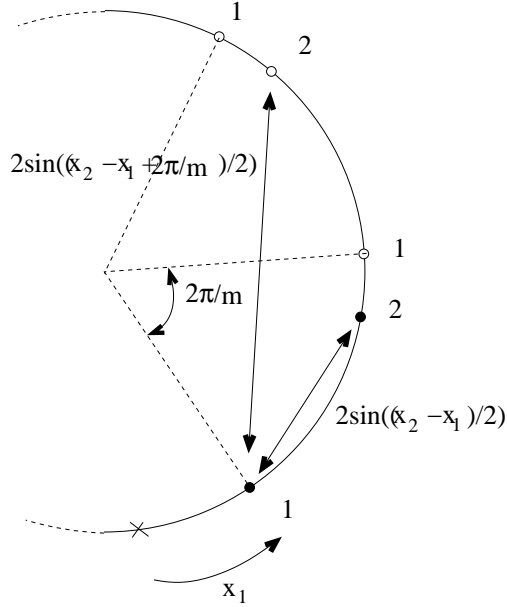,height=8cm}
\caption{Partial picture of the model for $N=2$. The particles are represented 
by full circles whereas their images by rotations of $2\pi/m$ are represented by empty circles.
\label{pic2}}
\end{center}
\end{figure}

One sees that each particle $x_i$ interacts with every other particle $x_j$ both directly and via
its images under rotations. A sketch of the case $G(3,1,2)$ is shown in figure \ref{pic2}.   Note
that the Hamiltonian is not local, because of the final term which exchanges (and moves) particles.
To find local Hamiltonians it is useful to introduce spins, as follows.  

\section{Particles with spin \label{sec:spin}}

We now generalize the models above to particles with internal `spin' degrees of freedom, $\vec s \in
\mathbb C^n$. Let us introduce a map $Q\in U(n)$ of order $m$ ($Q^m=1$) and write $Q_i$ for $Q$
acting on the spin $\vec s_i$ of the $i^{th}$ particle. Let also $P_{ij}$ be the operator which
exchanges the spins of the the $i^{th}$ and $j^{th}$ particles. $P_{ij}$ and $Q_i$ then obey the
same
defining relations of $G(m,1,N)$ as $\P_{ij},\Q_i$ in (\ref{Gm1N}). The two copies of $G(m,1,N)$
commute.  

The introduction of spins typically makes the restriction to the case of identical particles much
richer and more interesting. For bosons (fermions) the wavefunction
should now be (anti)symmetric under exchange of positions \emph{and} spins -- that is, under
\be \P_{ij} P_{ij}.\ee 
These generate the group $S_N$ of exchange symmetries. In the original $A_N$-series Sutherland
models, on wavefunctions with definite exchange statistics it is possible  \cite{Poly2,BGHP}
systematically to eliminate $\P_{ij}$ in favour of $P_{ij}$ in the Hamiltonian and higher conserved
quantities, and so obtain a purely local model with spin-spin interactions. 

We would like to do something similar in the present case. Here the exchange-symmetry group is
contained in several larger groups of discrete symmetries (involving the $Q_i$ and $\Q_i$). It is
natural to pick one of these as a group of ``generalized exchange symmetries'' and demand definite
behaviour of the wavefunction under it. There are a number of possibilities: one could for example
 demand that a full copy of $G(m,1,N)$, generated by e.g. $ \P_{ij} P_{ij}$ and $\Q_i Q_i$,
be promoted in this sense. But to do so would be overly restrictive on physical wavefunctions;
instead, it will suffice to demand invariance under
\be \Q^{-s}_i \P_{ij} \Q^s_i \,  Q^{-s}_i P_{ij} Q^s_i \label{qpq}\ee   
for all $i\neq j$ and for all $s=0,1,\dots,m-1$. These generate a copy of $G(m,m,N)$. 

(We focus for definiteness on $+1$-eigenstates of (\ref{qpq}), but the more general case with
arbitrary eigenvalues $p_s\in\{\pm 1\}$ for each value of $s$ -- in particular, $p_0=-1$, giving
fermions -- can be treated very similarly.) 

Let $\Lambda$ be the projector onto such states. To write $\Lambda$ explicitly, let $g\mapsto P_g$
and $g\mapsto \P_g$ be the maps representing abstract elements $g\in G(m,m,N)$ on, respectively,
spins and positions. Then  
\be \label{defL}
\Lambda = \frac{1}{N!\, m^{N-1}} \sum_{g\in G(m,m,N)} P_g \mathscr \P_g\ee
and we consider wavefunctions such that
\be \psi = \Lambda \psi.\label{proj}\ee
Define $\Is^{(k)}$ to be the operator obtained by first moving all the $\P_{ij}$, $\Q_i$ in
$I^{(n)}$, as defined in (\ref{Hi}), to the right of all positions $x_i$ and derivatives
$\frac{\del}{\del x_i}$, and then replacing them $\P_{ij} \mapsto P_{ij}$, $\Q_i \mapsto
Q_i$.\footnote{The replacement map could also be defined, following \cite{BGHP}, as the projection
\be\pi : G(m,1,N)_{\left<\P P, \Q Q^{-1}\right>} \ltimes G(m,1,N)_{\left<P,Q\right>} \rightarrow
G(m,1,N)_{\left<P,Q\right>}\,\,;\,\, A B \mapsto B.\ee Then indeed $\pi \P_{ij} = \pi (\P_{ij}
P_{ij}) P_{ij} = P_{ij}$ and $\pi \Q_i = \pi (\Q_i Q^{-1}_i) Q_i = Q_i$.}
It follows from the property 
\be \left(\Q_i^{-s} \P_{ij}\Q_i^s\right) \Lambda = \left(Q_i^{-s} P_{ij} Q_i^s\right)
\Lambda,\label{lprps}\ee
of $\Lambda$ that
\be \Is^{(k)} \Lambda = I^{(k)} \Lambda \label{sol}\ee
so that these operators agree on wavefunctions obeying (\ref{proj}). 

Note next the following properties of the $I^{(k)}$:
\be \left[ I^{(k)}, \P_{ij} \right] = 0,\quad \left[ I^{(k)}, \Q_i \right] = 0\ee
which follow from the definition $I^{(k)} = \sum_{i=1}^N d_i^k$ and the algebra
(\ref{dprops}--\ref{dprops2}) of
the Dunkl operators. It is also trivially the case that
\be \left[ I^{(k)}, P_{ij} \right] = 0,\quad \left[ I^{(k)}, Q_i \right] = 0.\ee
Consequently, for any monomial $M(\{I^{(k)}\})$ in the $I^{(k)}$, $M(\{I^{(k)}\})\Lambda$ obeys the
same relations (\ref{lprps}) as $\Lambda$ itself,
\be \left(\Q_i^{-s} \P_{ij}\Q_i^s \right)M(\{I^{(k)}\})\Lambda = \left(Q_i^{-s} P_{ij} Q_i^s\right)
M(\{I^{(k)}\}) \Lambda, \label{Mlprops}\ee
and thus
\be \Is^{(k)} M(\{I^{(p)}\})\Lambda = I^{(k)}M(\{I^{(p)}\}) \Lambda \label{sMol}.\ee
Given now any string of $\Is^{(k)}$'s, not \emph{a priori} assumed to commute, repeated use of this
fact allows one to replace each $\Is^{(k)}$ by $I^{(k)}$, working from the inside out:
\be \left(\dots \Is^{(k)} \Is^{(\ell)}\right) \Lambda = \left(\dots \Is^{(k)} I^{(\ell)}\right)
\Lambda = \left( \dots I^{(k)} I^{(\ell)}\right) \Lambda = \dots.\ee  
Having done so, the result
\be \left[ I^{(k)}, I^{(\ell)}\right] = 0,\ee
may be used to reorder the $I^{(k)}$ at will, and the above procedure then reversed to return
$I^{(k)} \rightarrow \Is^{(k)}$. Thus, in particular, we have that the $N$ independent evolution
operators 
\be U_k(t)= e^{it \Is^{(k)}}, \quad k=1,\dots,N\ee
commute amongst themselves when acting on physical wavefunctions:
\be U_k(t) U_\ell(t') \Lambda =   U_\ell(t') U_k(t) \Lambda.\ee
Therefore the model described by the Hamiltonian
\be
H_{spin}=\Is^{(2)} = -\sum_{i=1}^N \frac{\del^2}{\del x_i^2}
+\frac{\lambda}{4} \sum_{i\neq j}\sum_{s\in\mathbb Z_m}\
\frac{1}{\sin^2\left(\half\left(x_i-x_j+{2\pi s\over m}\right)\right)}
\left(\lambda+Q_i^{-s}P_{ij}Q_i^s\right)
\ee
is integrable.

\section{Static spin chain}\label{sec:static}

In this section, we find static integrable spin models from the new integrable models introduced
above. Indeed, it is well-known that from the $A_N$ Sutherland model it is possible to find static
spin chains \cite{Poly3,Ber}, called usually Haldane--Shastry models \cite{HS1,HS2}. Using similar
methods, we will prove that the following Hamiltonian
\bea
\label{Hamstat}
\overline H =\sum_{i\neq j}\sum_{s\in \Z_m}
\frac{\tau^s\ q_iq_j}{(q_i-\tau^s q_j)^2}
\ \Q_i^{-s}\P_{ij}\Q_i^s
\eea
is integrable for some particular values of the positions $q_i$.\\
First, we introduce the following operators
\bea
\overline d_i&=&\sum_{j< i}\sum_{s\in\Z_m}\frac{q_i}{q_i-\tau^s q_j}
\ \Q_i^{-s}\P_{ij}\Q_i^s
+\sum_{j> i}\sum_{s\in \Z_m}\frac{\tau^s q_j}{q_i-\tau^s q_j}
\  \Q_i^{-s}\P_{ij}\Q_i^s\qquad
\eea
Note that $d_i=q_i\frac{\partial}{\partial q_i}+\lambda \overline d_i$. Since the relation
$[d_j,d_k]=0$ is valid for any $\lambda$, we deduce that
\bea
[\overline d_j,\overline d_k]=0\; .
\eea
Similarly, from the relation $[H,d_i]=0$, it follows that, for any $i$,
\bea
[\overline H, \overline d_i]=\left[q_ip_i\ ,\ 
\sum_{j\neq\ell}\sum_{s\in\Z_m}\frac{\tau^s\ q_jq_\ell}{(q_l-\tau^sq_j)^2}\right]
\eea
The commutators $[\overline H, \overline d_i]$ therefore vanish if and only if
\bea
\sum_{j\neq i}\sum_{s\in\Z_m}\tau^s 
\frac{q_iq_j(q_i+\tau^s q_j)}
{(q_i-\tau^sq_j)^3}=0\quad,\forall i=1,2,\dots,N
\eea
and these conditions are fulfilled if 
\be\label{qqk}
q_k=\exp\left(\frac{2ik}{mN}\right)\ .\ee
Then, the Hamiltonian (\ref{Hamstat}) with the particular values of the positions given by
(\ref{qqk}) is integrable and may be written as follows
\bea
\overline H =-\frac{1}{4}\sum_{k\neq \ell}\sum_{s\in\Z_m}
\frac{1}{\sin^2\left(\frac{\pi}{m N}(k-\ell-Ns)\right)}
\ \Q_k^{-s}\P_{k\ell}\Q_k^s\;.
\eea 
By the same procedure as in the section \ref{sec:spin}, we can obtain an integrable
Hamiltonian acting on spins
\bea
\overline H_{spin} =-\frac{1}{4}\sum_{k\neq \ell}\sum_{s\in \Z_m}
\frac{1}{\sin^2\left(\frac{\pi}{m N}(k-\ell-Ns)\right)}
\ Q_k^{-s} P_{k\ell} Q_k^s  
\eea 

\begin{figure}[ht]
\begin{center}
\epsfig{file=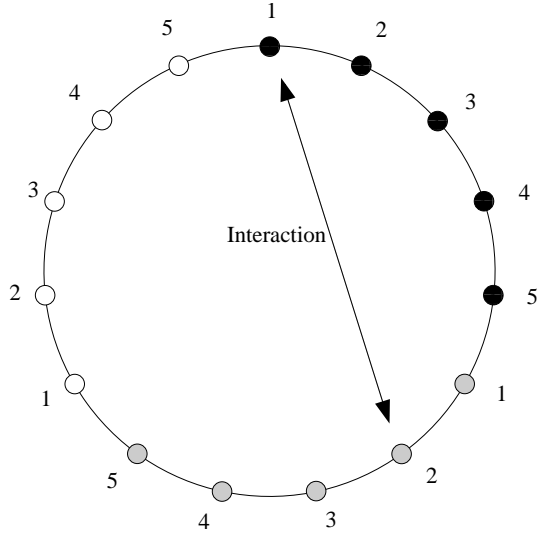,height=7cm}
\caption{Position of the spins on a circle of diameter 1 for $N=5$ and $m=3$.\label{pic}}
\end{center}
\end{figure}

\section{Models with dihedral symmetry}\label{dihed}

In the previous sections we introduced Sutherland models based on the 
complex reflection group $G(m,1,N)$.
They reduce to the original Sutherland models, as in (\ref{S1}), in the special case $G(1,1,N)=
A_{N-1}$, but for other values of $m$ they are new. In particular, although there is an isomorphism
of groups
\be G(2,1,N) \cong BC_N \cong \left(\mathbb Z_2\right)^N \rtimes S_N, \ee
as we noted above, our models certainly do not coincide with the usual $BC_N$ Sutherland models in
this case, because the $\mathbb Z_2$ generator is realized in different ways. In the $G(m,1,N)$
models, recall,
\be \Q_{i} \psi(\dots,q_i,\dots) = \psi(\dots,\tau q_{i},\dots) ,
\ee 
where $\tau= e^{2\pi i/m}$, whereas in the $BCD_N$ case the action of the $\mathbb Z_2$ generator is
\cite{Ber},
\be \K_{i} \psi(\dots,q_i,\dots) = \psi(\dots,q^{-1}_{i},\dots)\;. \label{Kaction}\ee 
In this section we show that it is possible to include both types of symmetry, rotation and
reflection.

\subsubsection*{Dunkl operators for wreath products of dihedral groups}

To take into account the new operators $\K_i$, we must find the group $W(m,N)$ generated by the
$\Q_i$, $\K_i$ and $\P_{ij}$, which will contain as subgroups both $D_N$ and $G(m,1,N)$. First note
that $\Q_i$ and $\K_i$ satisfy, for each $i$, 
\be \Q_i^m=1\quad,\quad\K_i^2=1\quad\text{and}\quad \K_i \Q_i =(\Q_i)^{-1}\K_i \ , \label{qk}\ee
which are the defining relations of the dihedral group of order $m$, denoted $\Dih_m$. We deduce
that the group $W(m,N)$ must be the wreath product
\be W(m,N)=\Dih_m \wr S_N  = (\Dih_m)^N \rtimes S_N.\ee
A minimal set of generators for $W(m,N)$ is 
\be\{\q , \k, \e_1, \e_2,\dots, \e_{N-1}\},\ee 
obeying the relations
\bea \e_i^2 = 1\quad & \quad \e_i \e_{i+1} \e_i = \e_{i+1} \e_i \e_{i+1}\quad& 
                     \quad \e_i \e_j = \e_j \e_i \quad (|i-j|>2) \nn\\
   \q ^m = 1 \quad& \quad \q  \e_1 \q  \e_1 = \e_1 \q  \e_1 \q  \quad& \quad \q \e_j = \e_j \q 
\quad (j>1) \nn \\\label{dih}
\k \q  =\q ^{-1}\k\qquad \k^2=1 \quad&\quad \k \e_{1} \k \e_{1}=\e_{1} \k \e_{1} \k \quad& \quad \k
\e_j=\e_j \k \quad(j>1)
\eea
and in terms of these $\P_{ij}$ and $\Q_i$ are again defined as in (\ref{Pdef}) and (\ref{Qdef})
while 
\bea \K_1 &=& \k \\\K_i &=& \P_{i1} \K_1 \P_{i1} \quad (i>1).\label{Kdef}\eea

To construct integrable models, we must extend this algebra as explained in the previous sections.
Define $H_{\lambda,\mu}(W(m,N))$, $\lambda, \mu\in \mathbb C$, to be the algebra generated by
$\{\dd, \k, \q, \e_1, \e_2,\dots, \e_{N-1}\},$ obeying (\ref{dih}) and the further relations
\be \k \dd = -\dd \k + \mu \sum_{s\in \Z_m} \q^{2s} \qquad,\qquad  \q  \dd=\dd \q\label{drels1D} \ee
\be \left(\dd + \lambda\sum_{s\inZm} \q^{-s} \e_1 \q^s \right)  \e_1 \k \e_1 = \e_1 \k
\e_1 \left(\dd + \lambda\sum_{s\inZm} \q^{-s} \e_1 \q^s \right)  \quad,\quad \dd \e_1 \q 
\e_1 = \e_1 \q  \e_1 \dd  \label{dst}\ee
\be \dd \left(\e_1 \dd \e_1 + \lambda \sum_{s\inZm} \q ^{s} \e_1 \q ^{-s}\right) =
\left(\e_1 \dd \e_1 + \lambda \sum_{s\inZm} \q ^{s} \e_1 \q ^{-s}\right) \dd \quad,\quad
\e_j \dd = \dd \e_j \quad (j>1)\label{drelsD}.\ee
Defining as before
\bea \dd_1 &=& \dd\nn\\ \dd_{i+1} &=& \P_{ii+1} \dd_i \P_{ii+1} 
 + \lambda \sum_{s \inZm}\Q_i^{s} \P_{ii+1} \Q_i^{-s} 
\quad (i=1,\dots,N-1)\label{dpropsD}\eea
it follows from (\ref{drels1D}-\ref{drelsD}) that
\be \left[ \,\dd_i, \dd_j \right] = 0, \quad \quad \left[\,\dd_i, \Q_j\right] = 0.\ee

The hard step, just in the case of $G(m,1,N)$, is to find a concrete realization of the $\dd_i$
satisfying these abstract relations. 
\begin{theorem}
For any $\rho \in \mathbb C$, the differential operators
\bea \dd_i &=& q_i \frac{\del}{\del q_i} \label{bdunkl}
  + \lambda \sum_{j\neq i}\sum_{s\in\mathbb Z_m} 
         \left( \frac{q_i}{q_i-\tau^s         q_j} \Q_i^{-s}\P_{ij}\Q_i^s
               +\frac{q_i}{q_i-\tau^{-s}  q^{-1}_j} \K_i \Q_i^{-s}\P_{ij}\Q_i^s \K_i \right)\nn\\
&& {} + \sum_{s\in\mathbb Z_m} \frac{\mu\tau^sq_i-\rho}{\tau^{s}q_i-\tau^{-s} q^{-1}_i}
        \Q_i^{2s}\ \K_i \nn\\
&& {}-\lambda\sum_{j>i}\sum_{s\in\mathbb Z_m}\Q_i^{-s}\P_{ij}\Q_i^s\label{newD}\eea
obey (\ref{drels1D}-\ref{drelsD}) and (\ref{dpropsD}).
\end{theorem}

We have verified this by direct computation, which is conceptually straightforward though somewhat
laborious. It is also possible to adapt the proof we used in the $G(m,1,N)$ case above, as follows.

\proof
Some details of the verification of (\ref{drels1D}) are given in an appendix, but as in the
$G(m,1,N)$ case the important and difficult step is to show that 
\be [\dd_i,\dd_j] = 0.\label{ddC}\ee
The Dunkl operators of the $BC_N$ case \cite{caduk} may be written
\bea Y_i &=& q_i\frac{\del}{\del q_i} + m\lambda\sum_{j\neq i}\left( \frac{q_i}{q_i-q_j} \P_{ij}
                                 +\frac{q_i}{q_i-q_j^{-1}} \K_i \P_{ij} \K_i\right)  \nn\\
                  &&{} + \frac{\mu q_i - \rho}{q_i - q_i^{-1}} \K_i - m\lambda\sum_{j>i}
\P_{ij}.\eea
It follows from the result 
\be \left[ Y_i, Y_j\right] = 0\ee
that, with the projectors $\Pi_i^r$ as defined in (\ref{Pri}), 
\be 0= \Pi_i^0 \Pi_j^0 \left[ Y_i , Y_j \right] =
    \sum_{r,t\inZm} \left[ \Pi_i^r \Pi_j^t Y_i, \Pi_i^{-r} \Pi_j^{-t} Y_j\right]\label{ppyy}.\ee
Now one can compute, for $i\neq j$,
\bea \Pi_i^r \Pi_j^t Y_i &=& \delta^{r,0} \delta^{t,0} q_i \frac{\del}{\del q_i} + 
 \delta^{t,0} \lambda \sum_{s\inZm} \tau^{rs} \left( \sum_{h\not\in\{i,j\}} \frac{q_i}{q_i-\tau^s
q_h} \Q_i^{-s} \P_{ih} \Q_i^s - \sum_{h>i, h\neq j} \Q_i^{-s} \P_{ih} \Q_i^s \right) \nn\\
&&{}+ \delta^{t,0} \lambda \sum_{s\inZm} \tau^{-rs} \left( 
           \sum_{h\not\in\{i,j\}} \frac{q_i}{q_i-\tau^{-s} q_h^{-1}} \K_i \Q_i^{-s} \P_{ih} \Q_i^s
\K_i
         +  \frac{\mu \tau^s q_i - \rho}{\tau^s q_i - \tau^{-s} q_i^{-1}} \Q^{2s} \K_i
\right)\label{ppy}\\
&&{} + \delta^{r+t,0}\lambda\sum_{s\inZm} \tau^{rs}\left( \frac{q_i}{q_i-\tau^s q_j} \Q_i^{-s}
\P_{ij} \Q_i^s + \frac{q_i}{q_i-\tau^{-s} q_j^{-1}} \K_i \Q_i^{-s} \P_{ij} \Q_i^s \K_i -
\theta^{j>i} \Q_i^{-s} \P_{ij} \Q_i^s \right).\nn\eea
The $\delta$'s mean that the only terms that can possibly survive in (\ref{ppyy}) are 
\bea 0&=& \left[ \Pi_i^0 \Pi_j^0 Y_i , \Pi_i^0 \Pi_j^0 Y_j \right] 
   + \sum_{t\inZm, t\neq 0}\left[\Pi_i^t \Pi_j^{-t} Y_i, \Pi_i^{-t} \Pi_j^t Y_j\right] .\eea
As before, the second term vanishes on closer inspection, and since  $\Pi_i^0 \Pi_j^0 Y_i=\dd_i$ by
(\ref{ppy}), we have established (\ref{ddC}) as required.
\finproof 

\subsubsection*{Integrable models}

Setting $\beta=(\mu+ \rho)/2$ and $\gamma=(\mu- \rho)/2$, we may rewrite the Dunkl operators
(\ref{bdunkl}) as
\bea
\label{bdunkl2}
D_\ell=d_\ell+\lambda \sum_{k\neq\ell}\sum_{s=0}^{m-1}\frac{\tau^s q_\ell\ q_k}{\tau^s q_\ell\
q_k-1}
 \K_\ell\Q_\ell^{-s} \P_{\ell k}\Q_\ell^s \K_\ell
+\sum_{s=0}^{m-1}\left(\frac{\beta\tau^sq_\ell}{\tau^{s}q_\ell+1}
+\frac{\gamma\tau^sq_\ell}{\tau^{s}q_\ell-1}\right)\Q_\ell^{2s}\ \K_\ell 
\eea
This equivalent form is useful for computing Hamiltonians. As explained
in the previous sections, we know that the model described by the Hamiltonian
$\cH=\sum_{\ell=1}^N D_\ell^2$ is integrable because it is one of a set of $N$ independent
mutually-commuting conserved quantities, namely
\be
J^{(k)}=\sum_{\ell=1}^N\ D_\ell^k\;.
\ee
The explicit form of the Hamiltonian depends on the parity of $m$: for $m$ odd, 
\bea
\cH^{odd}&=&H-\lambda\sum_{k\neq \ell}\sum_{s\in \Z_m}
\frac{\tau^s q_\ell q_k}{(\tau^sq_\ell q_k-1)^2}
(\lambda+\K_\ell\Q_\ell^{-s}\P_{\ell k}\Q_\ell^s\K_\ell)\nonumber\\
&&+\sum_{\ell}\sum_{s\in \Z_m}\left(\frac{\beta\tau^sq_\ell}{(1+\tau^sq_\ell)^2}(\beta
+\Q_\ell^{2s}\K_\ell)-\frac{\gamma\tau^sq_\ell}{(1-\tau^sq_\ell)^2}(\gamma
+\Q_\ell^{2s}\K_\ell)\right)
\eea
-- where $H$ is given by (\ref{ham}) -- while for $m$ even 
\bea
\cH^{even}&=&H-\lambda\sum_{k\neq \ell}\sum_{s\in \Z_m}
\frac{\tau^s q_\ell q_k}{(\tau^sq_\ell q_k-1)^2}
(\lambda+\K_\ell\Q_\ell^{-s}\P_{\ell k}\Q_\ell^s\K_\ell)\nonumber\\
&&-\mu\sum_{\ell}\sum_{s\in \Z_m}\frac{\tau^sq_\ell}{(1-\tau^sq_\ell)^2}(\mu +\Q_\ell^{2s}\K_\ell).
\eea
In the case $\beta=\gamma$ (i.e. $\rho=0$), the boundary term in the Hamiltonian, for $m$ odd, can
be simplified and becomes
\be
-\sum_{\ell}\sum_{s\in \Z_{2m}}\frac{\beta\sqrt{\tau^s}q_\ell}{(1-\sqrt{\tau^s}q_\ell)^2}(\beta
+\Q_\ell^{s}\K_\ell)
\ee  
After the change of coordinates $q_\ell=\exp(ix_\ell)$, the Hamiltonians are
\bea
\cH^{odd}&=&H+\frac{\lambda}{4}\sum_{k\neq \ell}\sum_{s\in \Z_m}
\frac{1}{\sin^2\left(\half\left(x_\ell+x_k+{2\pi s\over m}\right)\right)}
(\lambda+\K_\ell\Q_\ell^{-s}\P_{\ell k}\Q_\ell^s\K_\ell)\\
&&{}+\sum_{\ell}\sum_{s\in \Z_m}\left(
\frac{\beta/4}{\cos^2\left(\half\left(x_\ell+{2\pi s\over m}\right)\right)}
(\beta +\Q_\ell^{2s}\K_\ell)+
\frac{\gamma/4}{\sin^2\left(\half\left(x_\ell+{2\pi s\over m}\right)\right)}
(\gamma +\Q_\ell^{2s}\K_\ell)\right)\nonumber\\&&\nn \\
\cH^{even}&=&H+\frac{\lambda}{4}\sum_{k\neq \ell}\sum_{s\in \Z_m}
\frac{1}{\sin^2\left(\half\left(x_\ell+x_k+{2\pi s\over m}\right)\right)}
(\lambda+\K_\ell\Q_\ell^{-s}\P_{\ell k}\Q_\ell^s\K_\ell)\nonumber\\
&&{}+\frac{\mu}{4}\sum_{\ell}\sum_{s\in \Z_m}\frac{1}{\sin^2\left(\half\left(x_\ell+{2\pi s\over
m}\right)\right)}(\mu +\Q_\ell^{2s}\K_\ell)
\eea
where now $H$ is given by (\ref{ham-sin}).

\subsubsection*{Models on spins}

As explained in the section \ref{sec:spin}, it is possible to construct 
models acting on spins using the suitable projectors on the wavefunctions.
In addition to the map $Q$ introduced at the beginning of the 
section \ref{sec:spin}, we introduce now $K\in U(n)$ such that
\be
K^2=1\qquad\qquad K Q = Q^{-1} K\;.
\ee
Such matrices certainly exist: for example
\bea
Q=\text{diag}(\tau^{a_1},\dots,\tau^{a_n})\quad\text{with}\quad a_i=-a_{n+1-i}\quad\text{and}\quad
K=\text{antidiag}(1,\dots,1)\;.
\eea
In addition to the condtions (\ref{qpq}) on the wavefunctions, we demand that the physical wave
functions be invariant under
\be
\Q_i^{2s}\K_i\ Q_i^{2s}K_i \qquad\qquad\forall i,s
\ee
The explicit form of the projector is the product $\Lambda\Lambda_b$ where $\Lambda$ is defined by
(\ref{defL}) and
\be
\Lambda_b=\frac{1}{(2s)^N}\prod_{j}\left(\sum_{s\in \Z_m}
\Q_j^{2s} Q_j^{2s}\right)(1+\K_jK_j) \; .
\ee
At this point, a supplementary difficulty appears in comparison to the previous case (the same
problem appears in the usual $BC_N$ case in comparison to the $A_N$ case) because we get
$[J^{(k)},\Q_i]= 0$ and $[J^{(k)},\P_{ij}]= 0$ but
\be
[J^{(k)},\K_i]\neq 0.
\ee
Fortunately, we can show that this commutator vanishes when $k$ is even and, 
in particular, for $k=2$ which corresponds to the Hamiltonian.
Up to this restriction, we can use the same procedure to the section \ref{sec:spin} 
with the projector $\Lambda\Lambda_b$ and deduce 
that the dynamical spin model described by the Hamiltonian, for m even,
\bea
\cH_{spin}^{even}&=&H_{spin}+\frac{\lambda}{4}\sum_{k\neq \ell}\sum_{s\in \Z_m}
\frac{1}{\sin^2\left(\half\left(x_\ell+x_k+{2\pi s\over m}\right)\right)}
(\lambda+K_\ell Q_\ell^{-s}P_{\ell k}Q_\ell^s K_\ell)\nonumber\\
&&+\frac{\mu}{4}\sum_{\ell}\sum_{s\in \Z_m}\frac{1}{\sin^2\left(\half\left(x_\ell+{2\pi s\over
m}\right)\right)}(\mu +Q_\ell^{2s}K_\ell)
\eea
or, for m odd,
\bea
\cH^{odd}_{spin}&=&H_{spin}+\frac{\lambda}{4}\sum_{k\neq \ell}\sum_{s\in \Z_m}
\frac{1}{\sin^2\left(\half\left(x_\ell+x_k+{2\pi s\over m}\right)\right)}
(\lambda+K_\ell Q_\ell^{-s}P_{\ell k}Q_\ell^s K_\ell)\\
&&+\sum_{\ell}\sum_{s\in \Z_m}\left(
\frac{\beta/4}{\cos^2\left(\half\left(x_\ell+{2\pi s\over m}\right)\right)}
(\beta +Q_\ell^{2s} K_\ell)+
\frac{\gamma/4}{\sin^2\left(\half\left(x_\ell+{2\pi s\over m}\right)\right)}
(\gamma +Q_\ell^{2s} K_\ell)\right)\nonumber
\eea
is integrable.

\subsubsection*{Spin chain}

As explained in the section \ref{sec:static}, it is possible find an integrable static spin 
chain from a dynamical one. Using this procedure\footnote{It is convenient to rescale the coupling
constants of the boundary:
$\beta\rightarrow \lambda\beta$ and $\gamma\rightarrow\lambda\gamma$}, we can prove that the
following Hamiltonian, for m odd, 
\bea
\overline{\cH}^{odd}&=&\sum_{k\neq \ell}\sum_{s\in \Z_m}\left(
\frac{\tau^s q_\ell q_k}{(q_k-\tau^sq_\ell )^2}
\Q_\ell^{-s}\P_{\ell k}\Q_\ell^s\ +\ 
\frac{\tau^s q_\ell q_k}{(\tau^sq_\ell q_k-1)^2}
\K_\ell\Q_\ell^{-s}\P_{\ell k}\Q_\ell^s\K_\ell\right)\nonumber\\
&&+\sum_{\ell}\sum_{s\in \Z_m}\left(
\frac{\gamma\tau^sq_\ell}{(1-\tau^sq_\ell)^2}
-\frac{\beta\tau^sq_\ell}{(1+\tau^sq_\ell)^2}
\right)\Q_\ell^{2s}\K_\ell
\eea
is integrable if, for $\ell=1,\dots,N$,
\bea
\label{cond1}
\sum_{s\in \Z_m}\tau^s
\left(
2\sum_{j\neq\ell}\left(\frac{q_j(q_\ell+\tau^s q_j)}{(q_\ell-\tau^s q_j)^3}
+\frac{q_j(\tau^s q_\ell q_j+1)}{(\tau^s q_\ell q_j-1)^3}\right)+
\beta^2\frac{1-\tau^s q_\ell}{(1+\tau^s q_\ell)^3}-\gamma^2\frac{1+\tau^s q_\ell}{(1-\tau^s
q_\ell)^3}
\right) =0~~~
\eea
Similarly, for m even, we prove that the Hamiltonian
\bea
\overline{\cH}^{even}&=&\sum_{k\neq \ell}\sum_{s\in \Z_m}\left(
\frac{\tau^s q_\ell q_k}{(q_k-\tau^sq_\ell )^2}
\Q_\ell^{-s}\P_{\ell k}\Q_\ell^s\ +\ 
\frac{\tau^s q_\ell q_k}{(\tau^sq_\ell q_k-1)^2}
\K_\ell\Q_\ell^{-s}\P_{\ell k}\Q_\ell^s\K_\ell\right)\nonumber\\
&&+\sum_{\ell}\sum_{s\in \Z_m}
\frac{\mu\tau^sq_\ell}{(1-\tau^sq_\ell)^2}
\Q_\ell^{2s}\K_\ell
\eea
is integrable if, for $\ell=1,\dots,N$,
\bea
\label{cond2}
\sum_{s\in \Z_m}\tau^s
\left(
2\sum_{j\neq\ell}\left(\frac{q_j(q_\ell+\tau^s q_j)}{(q_\ell-\tau^s q_j)^3}
+\frac{q_j(\tau^s q_\ell q_j+1)}{(\tau^s q_\ell q_j-1)^3}\right)
-\mu^2\frac{1+\tau^s q_\ell}{(1-\tau^s q_\ell)^3}
\right) =0~~~
\eea
As discussed above, we can now replaced in the Hamiltonians $\overline{\cH}^{odd}$ 
and $\overline{\cH}^{even}$ the operators acting on the positions 
by the operators acting on spins, while preserved integrability.
We finish this section by discussing the different solutions of relations (\ref{cond1}) and
(\ref{cond2}) in which the positions are equidistant.
The solutions depend on the value of the coupling constant $\beta$ and $\gamma$ (or $\mu$ and
$\rho$).
Let us define $L$ to be the number of sites -- which may differ from $N$, the number of spins -- and
let 
$\omega_L=e^{2i\pi/L}$. Different possible distributions of the coordinates $q_i$, for m odd, are
given in the following table:
\begin{center}
\begin{tabular}{|c|c|c|c|c|}
\hline
$L$&$\beta^2$&$\gamma^2$&$q_k$&Figure\\[10pt]
\hline\hline
$2Nm$&$\displaystyle\frac{1}{4}$&$\displaystyle\frac{1}{4}$&
$\displaystyle\omega_L^{k-\half}$&\ref{pic3}\\[10pt]
\hline
{$2Nm+m$}
&$\displaystyle\frac{9}{4}$&$\displaystyle\frac{1}{4}$&$\displaystyle\omega_L^{k-\half}$&\ref{pic4}\\[10pt]
\cline{2-5}
&$\displaystyle\frac{1}{4}$&$\displaystyle\frac{9}{4}$
&$\displaystyle\omega_L^k$&\ref{pic6}\\[10pt]
\hline
$2(N+1)m$&$\displaystyle\frac{9}{4}$&$\displaystyle\frac{9}{4}$&$\displaystyle\omega_L^k$&\ref{pic5}\\[10pt]
\hline
\end{tabular}
\end{center}
The number in the column \textit{Figure} corresponds to the labels of the 
figures below where the particular case $m=3$ is taken. In these figures, the black circles
represent the positions of the original spins whereas the white circles represent the images of
these spins. The grey circles are empty sites.
Of course, we can recover the usual $BC_N$ cases studied in \cite{Ber} when we put $m=1$ in the
previous table.
The case when $m$ is even seems more complicated: we found no solution for  relation (\ref{cond2}).

\begin{figure}[htp]
\begin{minipage}[t]{7cm}
\epsfig{file=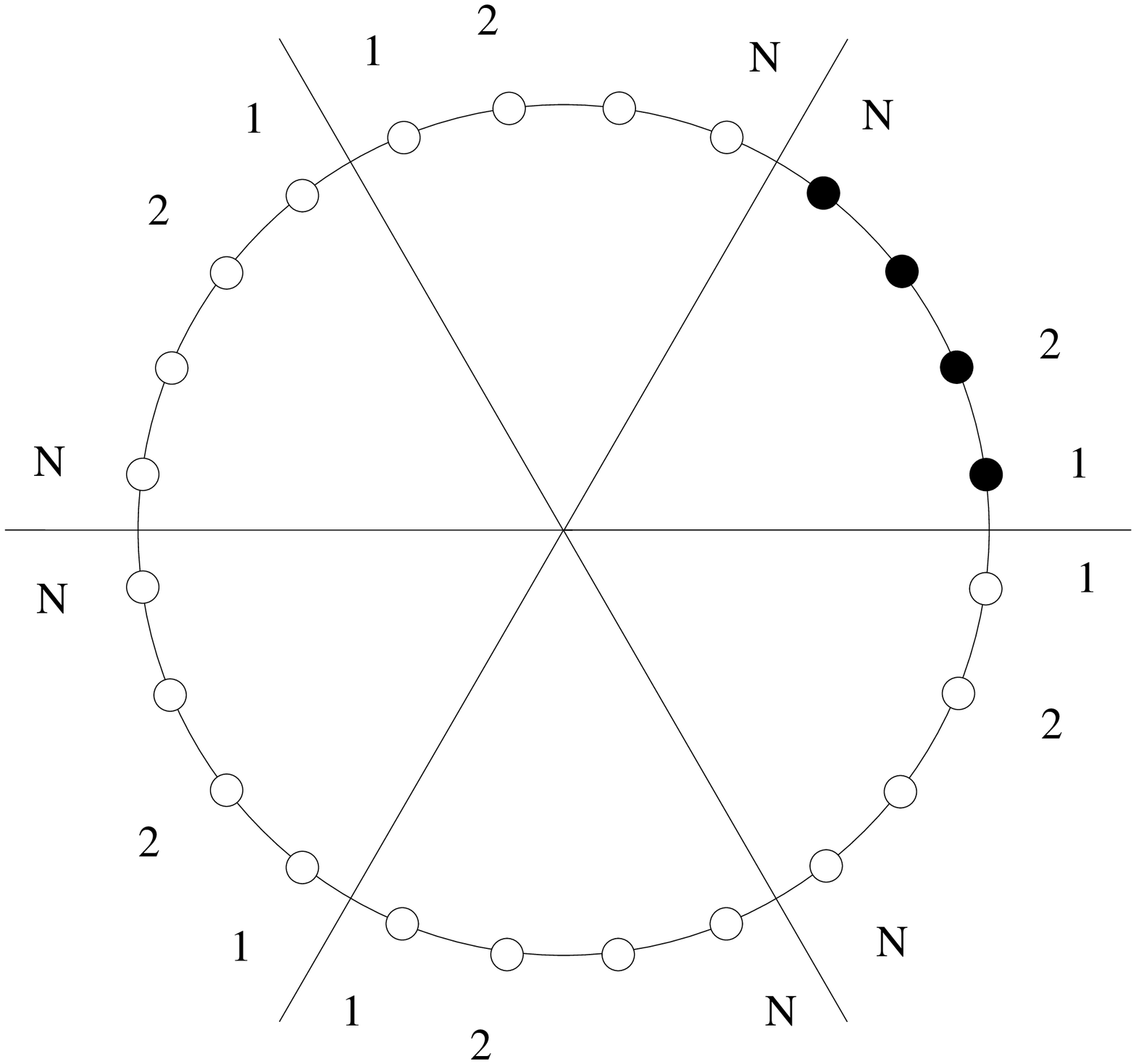,width=7cm}
\caption{\label{pic3}Position of the sites for $m=3$ and $L=2Nm$}
\end{minipage}
\hfill
\begin{minipage}[t]{7cm}
\epsfig{file=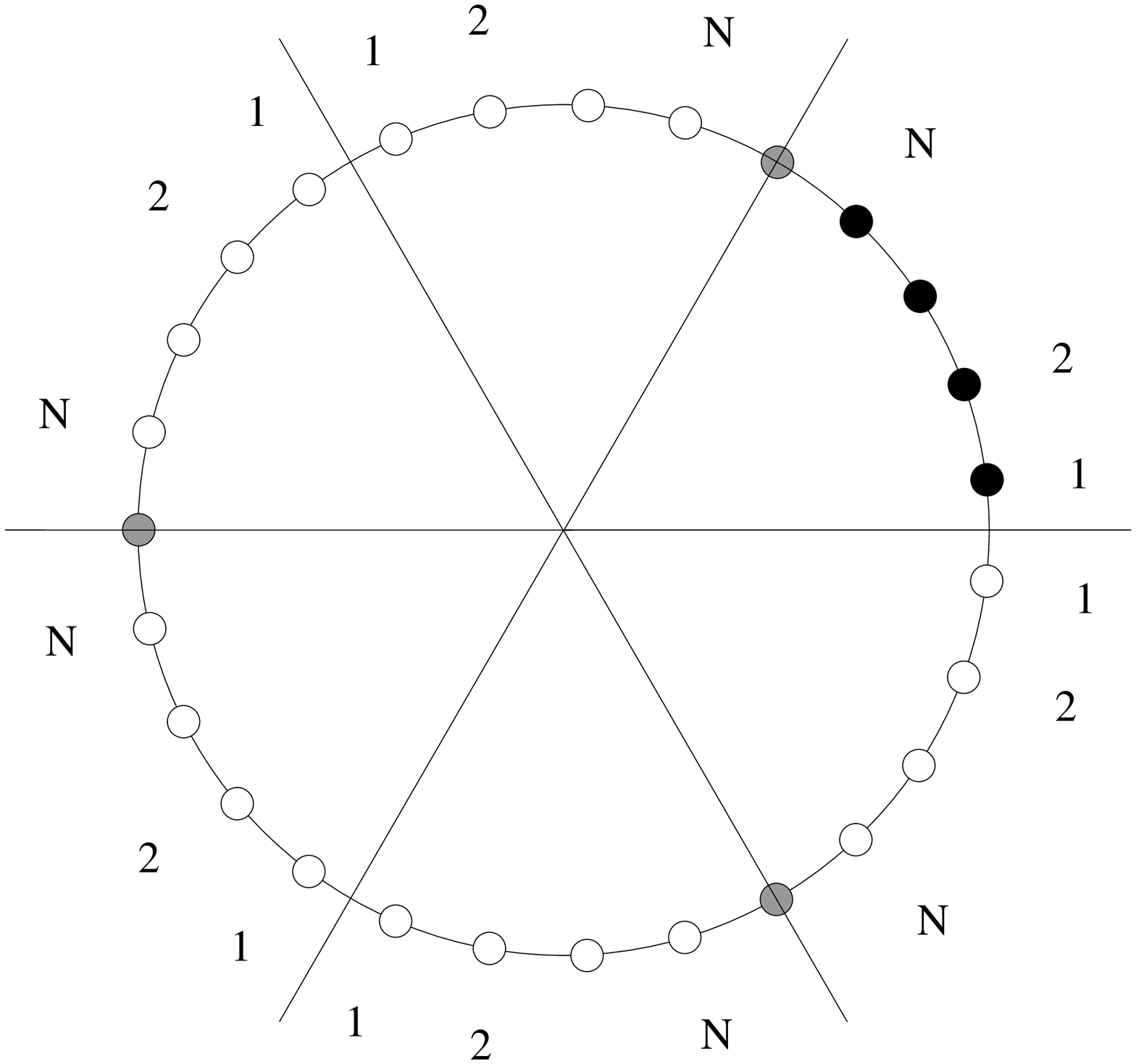,width=7cm}
\caption{\label{pic4} Position of the sites for $m=3$, $L=2Nm+m$ and $q_k=\omega_L^{k-\half}$}
\end{minipage}
\end{figure}

\begin{figure}[htp]
\begin{minipage}[t]{7cm}
\epsfig{file=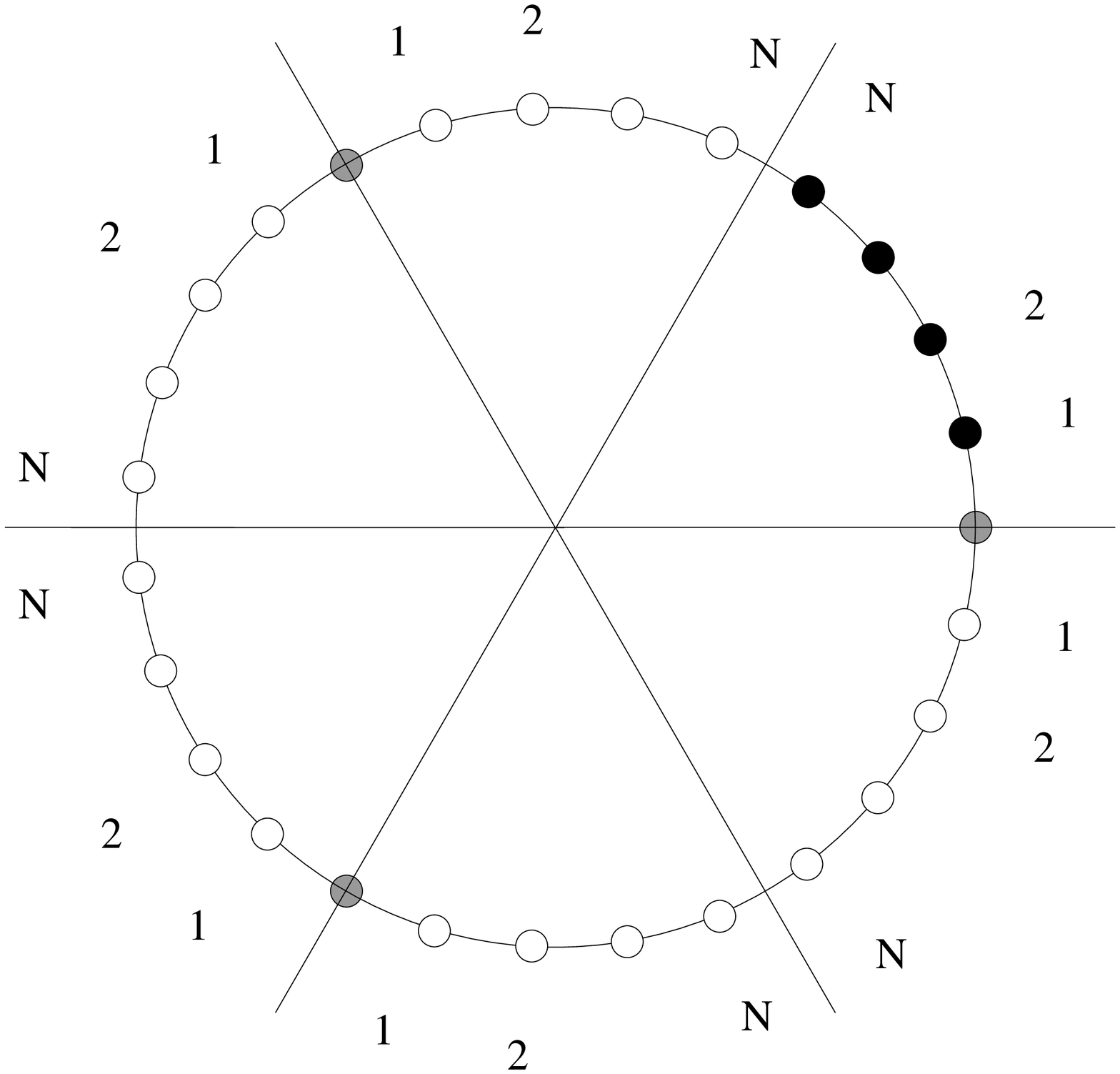,height=7cm}
\caption{\label{pic6} Position of the sites for $m=3$, $L=2Nm+m$ and $q_k=\omega_L^{k}$}
\end{minipage}
\hfill
\begin{minipage}[t]{7cm}
\epsfig{file=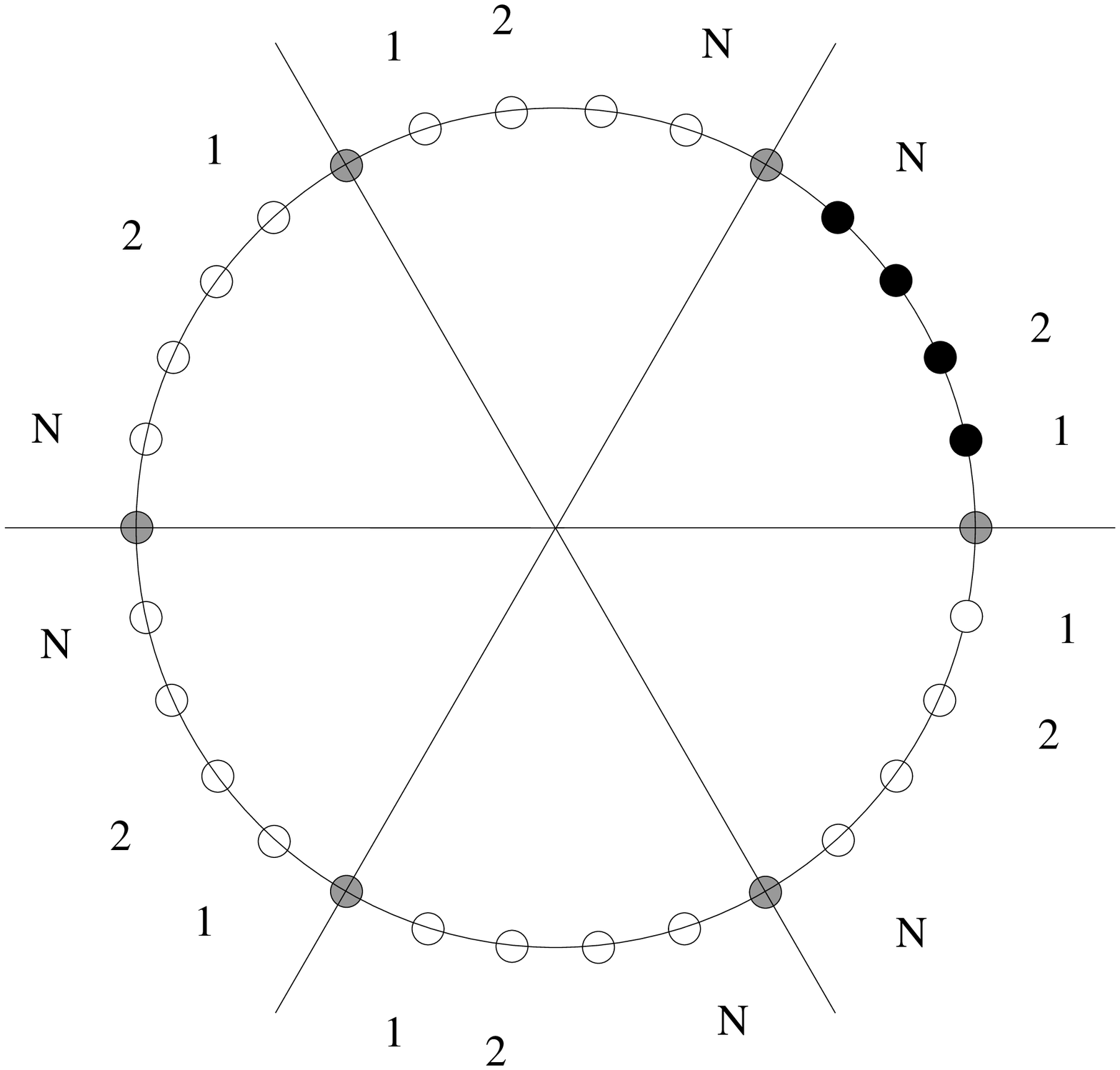,height=7cm}
\caption{Position of the sites for $m=3$ and $L=2(N+1)m$.
\label{pic5}}
\end{minipage}
\end{figure}

\section{Conclusions and outlook}
In this paper we considered two families of Sutherland models, in which each particle possesses
a set of images determined by a cyclic or dihedral group. The former we were led to by the desire to
find models in which complex reflection groups act as symmetries; in the latter, the role of the
reflection group is played by a wreath product of a dihedral group. The Dunkl operators were the key
ingredient in demonstrating integrability. In the cyclic case these had been found in \cite{duop},
but in the dihedral case they have not appeared previously, to the authors' knowledge.

We sought to emphasise the link between the models and complex reflection groups. In the cyclic cases the models themselves are a special case of systems previously obtained by appropriate reduction of a matrix model \cite{P1998nm} and of rational spin-Calogero models \cite{P1998gg}; in the latter case the equivalence may be seen by re-writing the $\sin^{-2}$ potential as an infinite sum of inverse squares. (Further generalizations of these models involving a ``twisted'' symmetry element were found in \cite{P01}.) In principle our models with dihedral symmetry could also be obtained by reductions of rational models involving parity in addition to translation symmetry, though this has not been done explicitly. 

There are a number of interesting open questions concerning these models. First, one should be able
to solve for the energy eigenstates exactly. This could be achieved by simultaneously diagonalizing
the Dunkl operators by means of (suitably generalized) Jack polynomials -- see e.g. \cite{LV}. One
also strongly expects, looking at figure \ref{fig2}, that it should be possible to obtain all the
models here from the standard $A$-series Sutherland model via a suitable reduction procedure
\cite{rev,PolyRev,OPreps}, just as is true of the $BC_N$ case. This is usually related to folding of
Dynkin diagrams (see e.g. \cite{BST}); here we expect broader notion of folding will come into play,
and the projectors used in our proof of commutation the Dunkl operators (theorem \ref{th1}) seem
rather suggestive. 

We stress however that these models are of interest in their own right, regardless of their origin
via reduction. In particular they should possess some extended symmetry algebra, analogous to the
Yangian and reflection-algebra symmetries of, respectively, the $A$ and $BC$ Sutherland models, but
respecting the underlying complex reflection group. It is worth remarking here on intriguing hints
in the mathematics literature that, the lack of root systems and so on notwithstanding, certain
complex reflection groups are actually closely analogous to real crystallographic ones (i.e. Weyl
groups) with -- very loosely speaking -- the role of Lie algebras being played by objects called
``spetses'' \cite{spetses}. These remain somewhat mysterious, and one can speculate that the
machinery of integrable models (Hopf algebras, $R$-matrices and so on) might provide a helpful new
perspective, as it has in the past for  Lie algebras and their representation theory. 

Finally let us note a few more open questions. Do there exist Sutherland models for the exceptional
complex reflection groups, perhaps via reduction as in \cite{Fring:2005am}? The Sutherland model is
the trigonometric member of the Calogero-Moser family: can one generalize to elliptic potentials?
What models, presumably conformal field theories \cite{Cadoni:2000cz}, are obtained in the limit of
large $N$?

\bigskip
\noindent{\bf Acknowledgements}

C.A.S.Y. gratefully acknowledges the financial support of the Leverhulme trust. 

\appendix
\section{Properties of Dunkl operators}
Here we give some details of the argument that $\dd_1=\dd$ obeys (\ref{drels1D}-\ref{drelsD}).
Consider the relation in (\ref{drels1D}) involving $b=\K_1$. 
For the terms at order $\lambda$, one finds first that
\bea && \K_1 \sum_{j\neq 1}\sum_{s\in\mathbb Z_m} 
         \left( \frac{q_1}{q_1-\tau^s         q_j} \Q_1^{-s}\P_{1j}\Q_1^s
               +\frac{q_1}{q_1-\tau^{-s}  q^{-1}_j} \K_1 \Q_1^{-s}\P_{1j}\Q_1^s \K_1 \right)\\
&&{} + \sum_{j\neq 1}\sum_{s\in\mathbb Z_m} 
         \left( \frac{q_1}{q_1-\tau^s         q_j} \Q_1^{-s}\P_{1j}\Q_1^s
       +\frac{q_1}{q_1-\tau^{-s}  q^{-1}_j} \K_1 \Q_1^{-s}\P_{1j}\Q_1^s \K_1 \right) \K_1 \nn\\
&=& \sum_{j\neq 1}\sum_{s\in\mathbb Z_m} \left( \frac{q^{-1}_1}{q^{-1}_1-\tau^{s}
q_j}+\frac{q_1}{q_1-\tau^{-s}  q^{-1}_j} \right) \K_1 \Q_1^{-s}\P_{1j}\Q_1^s + \left(
\frac{q^{-1}_1}{q^{-1}_1-\tau^{-s}  q^{-1}_j} +\frac{q_1}{q_1-\tau^s         q_j}  \right)
\Q_1^{-s}\P_{1j}\Q_1^s \K_1 \nn\\
&=& \sum_{j\neq 1}\sum_{s\in\mathbb Z_m} \K_1 \Q_1^{-s}\P_{1j}\Q_1^s + \Q_1^{-s}\P_{1j}\Q_1^s \K_1
\nn\eea
which then is precisely cancelled by the contribution from the other order-$\lambda$ piece in
(\ref{newD}). Note of course that the algebra of $\K_i$ with $q_i$, implicit the action
(\ref{Kaction}) of $\K_i$ on wavefunctions, is $\K_i\ q_i = q^{-1}_i\ \K_i$ and
$\K_i\frac{\del}{\del
q_i} = \frac{\del}{\del q^{-1}_i}\ \K_i$.
Since also 
\be  \K_1 q_1 \frac{\del}{\del q_1} = -q_1 \frac{\del}{\del q_1} \K_i\ee
(because $q_i^{-1}=q_i^{-1}$) we have that $\K_1 \dd_1 = - \dd_1 \K_1$ up to the terms involving
$\mu$
and $\rho$. It is straightforward to verify that these give
\be \K_1 \dd_1 = - \dd_1 \K_1 + \mu \sum_{s \inZm} \Q_1^{2s} \ee
as claimed. The other relation in (\ref{drels1D}), $\Q_1 \dd_1 = \dd_1 \Q_1$ is almost immediate. 

Next consider (\ref{dst}): $\K_2=\e_1 \k \e_1$ commutes term by term with the right hand side of
\bea \dd_1 + \lambda\sum_{s\inZm} \q^{-s} \e_1 \q^s &=& q_1 \frac{\del}{\del q_1}
\label{dd1} 
  + \lambda \sum_{j\neq 1}\sum_{s\in\mathbb Z_m} 
         \left( \frac{q_1}{q_1-\tau^s         q_j} \Q_1^{-s}\P_{1j}\Q_1^s
               +\frac{q_1}{q_1-\tau^{-s}  q^{-1}_j} \K_1 \Q_1^{-s}\P_{1j}\Q_1^s \K_1 \right)\nn\\
&& {} + \sum_{s\in\mathbb Z_m} \frac{\mu\tau^sq_i-\rho}{\tau^{s}q_i-\tau^{-s} q^{-1}_i} \Q_1^{2s}\
\K_1
\nn\\
&& {}+ \lambda\sum_{j>2}\sum_{s\in\mathbb Z_m}\Q_i^{-s}\P_{ij}\Q_i^s \eea
except when $j=2$ in the sum: but there
\bea &&\K_2 \sum_{s\in\mathbb Z_m} 
         \left( \frac{q_1}{q_1-\tau^s     q_2} \Q_1^{-s}\P_{12}\Q_1^s
               +\frac{q_1}{q_1-\tau^{-s}  q^{-1}_2} \K_1 \Q_1^{-s}\P_{12}\Q_1^s \K_1 \right)\nn\\
  &=& \sum_{s\in\mathbb Z_m} 
         \left( \frac{q_1}{q_1-\tau^s     q^{-1}_2} \K_2 \Q_1^{-s}\P_{12}\Q_1^s \K_2
               +\frac{q_1}{q_1-\tau^{-s}  q_2} \K_2\K_1 \Q_1^{-s}\P_{12}\Q_1^s \K_1\K_2 \right)\K_2
\nn\\
&=& \sum_{s\in\mathbb Z_m} 
         \left( \frac{q_1}{q_1-\tau^s     q^{-1}_2} \K_1 \Q_1^{s}\P_{12}\Q_1^{-s} \K_1
               +\frac{q_1}{q_1-\tau^{-s}  q_2}    \Q_1^{s}\P_{12}\Q_1^{-s} \right)\K_2 \nn\eea
and, on renaming $s\rightarrow -s$, one sees that the two terms are merely exchanged. Thus indeed
$\K_2$ commutes with $\dd + \lambda\sum_{s\inZm} \q^{-s} \e_1 \q^s$. To show that $\Q_2
\dd_1 = \dd_1 \Q_2$ is straightforward.

\end{document}